\documentclass[conference]{IEEEtran}
\IEEEoverridecommandlockouts
\usepackage{hyperref} 
\usepackage{cite}
\usepackage{amsmath,amssymb,amsfonts}
\usepackage{algorithmic}
\usepackage{graphicx}
\usepackage{textcomp}
\usepackage{xcolor}
\usepackage{booktabs}

\usepackage[normalem]{ulem}
\usepackage{soul}
\def\BibTeX{{\rm B\kern-.05em{\sc i\kern-.025em b}\kern-.08em
    T\kern-.1667em\lower.7ex\hbox{E}\kern-.125emX}}
\begin{document}

\title{Universal Sound Separation with Self-Supervised Audio Masked Autoencoder\\
}

\author{\IEEEauthorblockN{Junqi Zhao\textsuperscript{1},
Xubo Liu\textsuperscript{1},
Jinzheng Zhao\textsuperscript{1}, 
Yi Yuan\textsuperscript{1},
Qiuqiang Kong\textsuperscript{2},
Mark D. Plumbley\textsuperscript{1},
 Wenwu Wang\textsuperscript{1}}
\vspace{\baselineskip}
\IEEEauthorblockA{\textsuperscript{1}Centre for Vision, Speech and Signal Processing (CVSSP), University of Surrey, UK}
\IEEEauthorblockA{\textsuperscript{2}The Chinese University of Hong Kong (CUHK)}

}

\maketitle

\begin{abstract}
Universal sound separation (USS) is a task of separating mixtures of arbitrary sound sources. Typically, universal separation models are trained from scratch in a supervised manner, using labeled data. Self-supervised learning (SSL) is an emerging deep learning approach that leverages unlabeled data to obtain task-agnostic representations, which can benefit many downstream tasks. In this paper, we propose integrating a self-supervised pre-trained model, namely the audio masked autoencoder (A-MAE), into a universal sound separation system to enhance its separation performance. We employ two strategies to utilize SSL embeddings: freezing or updating the parameters of A-MAE during fine-tuning. The SSL embeddings are concatenated with the short-time Fourier transform (STFT) to serve as input features for the separation model. We evaluate our methods on the AudioSet dataset, and the experimental results indicate that the proposed methods successfully enhance the separation performance of a state-of-the-art ResUNet-based USS model.
\end{abstract}


\section{Introduction}\label{Intro}
Computational auditory scene analysis \cite{brown1994computational} aims to equip machine listening systems with the capability to selectively perceive numerous distinct events in the surrounding environment. Audio source separation \cite{makino2018audio}, as a fundamental task in computational auditory scene analysis, has been studied for many years \cite{Wang2005}. It has numerous applications across various domains, including automatic speech recognition \cite{narayanan2014investigation}, music transcription \cite{plumbley2002automatic}, and sound monitoring \cite{li2021convtasnet}.

In monaural source separation, the task is to segregate individual source tracks from a single-channel sound mixture without relying on any spatial cues. Many previous studies have primarily focused on specific types of sounds, such as speech \cite{wang2018supervised} or music \cite{cano2018musical}, and are only able to separate a limited number of sound sources. A more challenging problem is to separate arbitrary sound sources from each other using a single model, known as universal sound separation (USS). 

For single-channel USS tasks, separating all sources from an audio mixture is extremely challenging. In practical scenarios, we are generally interested in one particular source. Therefore, sound separation models can be designed to have a single output. To empower the model to extract arbitrary sources, we can condition the system on a query embedding derived from reference recordings of the desired source. This type of task is referred to as query-based source separation (QSS) \cite{lee2019audio}. The query information can take the form
of different modalities, such as audio \cite{chen2022zero}, vision \cite{zhao2018sound}, or language \cite{liu2022separate, liu2023separate}. 

In recent years, self-supervised learning (SSL) approaches have advanced rapidly and become the predominant method for pre-training models. SSL alleviates the reliance of supervised learning on large amounts of labeled data while exhibiting great performance and generalization capabilities \cite{liu2021self}. A growing number of SSL methods have been successfully applied in the field of audio, including applications such as audio classification \cite{huang2022masked}, speaker recognition \cite{fan2020exploring}, and speech recognition \cite{baevski2020effectiveness}. However, all these methods focus on representation learning for classification-based tasks. On the other hand, the source separation problem requires the model to estimate continuous target source signals.

One SSL model that has been applied for source separation is WavLM \cite{chen2022wavlm} for speech separation \cite{song2023exploring}. WavLM is a pre-trained model based on HuBERT \cite{hsu2021hubert}, which includes a CNN encoder and a Transformer. Another model, Pac-HuBERT \cite{chen2023pac}, has been proposed for music source separation, where HuBERT was adapted into a time-frequency domain separation model, achieving better separation performance compared to the ResUNet \cite{kong2021decoupling}. These SSL models undergo pre-training on the mask prediction task, where the objective is to predict masked tokens from visible tokens. While the representations learned by these models can be applied to source separation tasks, it's important to note that pre-training is typically performed using specific datasets, such as speech or music data. Therefore, identifying a suitable self-supervised model pre-trained on general audio data and extending it to universal sound separation has yet to be explored. 

In this work, we propose to use a self-supervised pre-trained audio model, i.e., the audio masked autoencoder (A-MAE), to extract general audio representations for improving USS models. Specifically, during the USS training stage, we propose to either freeze or partially update the parameters of the A-MAE to obtain the SSL representations, which are then concatenated with the short-time Fourier transform (STFT) features as input. Based on this, the downstream separator predicts the mask of the desired source. We evaluate our methods on the AudioSet dataset. Experimental results indicate that our proposed methods enhance the separation performance of a state-of-the-art (SOTA) ResUNet-based USS model \cite{kong2023universal}.

This paper is organized as follows: Section \ref{Rw} introduces several prior works related to USS and their limitations. Section \ref{FM} introduces the framework of query-based USS with weakly labeled data and our proposed method. Section \ref{Exper} presents the dataset, experimental setup, and evaluation method. Section \ref{RD} reports and analyzes the experimental results, while Section \ref{Conclu} summarizes our work.

\section{Related Work}\label{Rw}

The USS task was initially introduced in \cite{kavalerov2019universal}, where the authors proposed utilizing the iterative improved time-dilated convolutional network (iTDCN++) to separate mixtures with known numbers of sound sources. Next, the conditional information about which sound classes are present is used to improve universal sound separation performance in \cite{tzinis2020improving}. In order to reduce the cost of annotating data, Wisdom et al. \cite{wisdom2020unsupervised} proposed the mixture invariant training (MixIT) approach, a purely unsupervised source separation paradigm for training single-channel sound separation models on a large amount of unlabeled, in-the-wild data. Then, they also \cite{wisdom2021s} introduced a time-domain convolutional network (TDCN++) capable of separating an unknown number of sources in a mixture. The free universal sound separation (FUSS) dataset they used only consists of 357 sound categories. 

A zero-shot universal source separator was put forward by Chen et al. \cite{chen2022zero}, capable of leveraging AudioSet data for training and supporting unseen sources. Nonetheless, they did not report the separation performance of the system when using average embedding as query conditions on AudioSet. Recently, Kong et al. \cite{kong2023universal} proposed a method for training a USS system using weakly labeled data, achieving SOTA separation performance across  the 527 sound classes in the AudioSet dataset. The method of Kong et al. \cite{kong2023universal} will serve as our baseline.

\begin{figure*}[!ht]
  \centering
  \includegraphics[width=1.7\columnwidth]{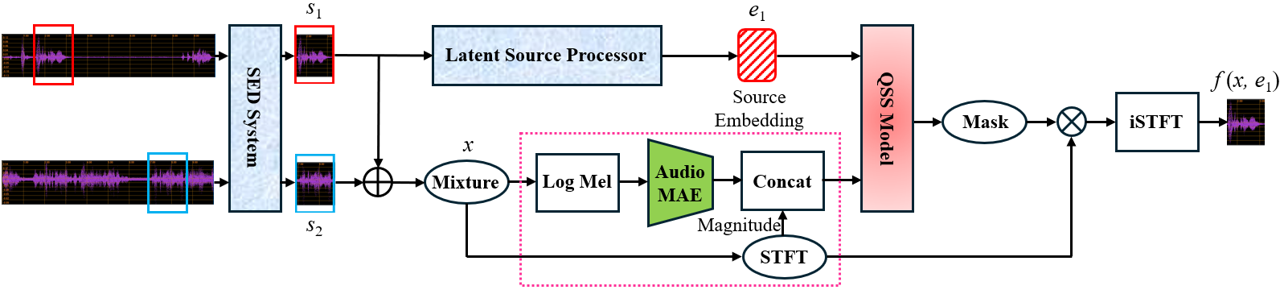}
  \caption{\textcolor{black}{The framework of our proposed query-based USS system.}}
  \label{framework}  
\end{figure*}

\section{Framework and Method}\label{FM}
\subsection{Query-based USS with Weakly Labeled Data}\label{QUSSWLD}
Most existing separation models require clean source and mixture pairs for training, and they can only separate a limited number of audio classes. To address this issue, Kong et al. proposed a method that utilizes a pre-trained sound event detection system to explore relatively clean target sound events \cite{kong2023universal}, making full use of the large-scale weakly labeled dataset AudioSet \cite{gemmeke2017audio}. As shown in Figure \ref{framework}, a typical query-based USS system with weakly labeled data comprises three components: a sound event detection (SED) system that localizes the occurrence of events in weakly labeled AudioSet training data; a query-based source separator trained on the refined data to separate an audio mixture into individual sources; and a latent source embedding (LSE) processor. The latent source processor controls the selection of sources to separate from a mixture, empowering the separator to segregate arbitrary sound sources.

\noindent \textbf{Sound Event Detection System.} The aim of the sound event detection task is to recognize the sound events and localize their occurrence (start and end) time in an audio recording. By leveraging a pre-trained SED model, we can utilize the frame-wise presence probability of the sound event to identify relatively clean sound sources from weakly labeled audio samples. The selected segment is referred to as the target anchor segment. Utilizing target anchor segments detected by the SED system as source data, mixtures can be constructed for training the USS system.

The SED model we employed is the Pretrained Audio Neural Networks (PANNs) \cite{kong2020panns}, which include VGG-like CNNs to convert an audio Mel-spectrogram into a $(T, K)$ feature map, where $T$ denotes the number of time frames, and $K$ denotes the number of sound classes. The feature map is a frame-wise prediction that indicates the probability of presence for each sound event at each time frame. We also utilize PANNs as our latent source embedding processor. To obtain the latent source embedding used in query-based sound separation, we average the output of the penultimate layer of PANNs along the time axis.

For a query-based USS system, we begin by randomly selecting two different audio samples from AudioSet. These samples are then fed into the SED system to extract target anchor segments for their respective classes. The 2-second target anchor segments for these two sound events are denoted as $s_1$ and $s_2$, respectively. Subsequently, these two segments are fed into the latent source processor to obtain two latent source embeddings, $e_1$ and $e_2$.

\noindent \textbf{Query-based Source Separator.} After obtaining $s_1$, $s_2$, $e_1$ and $e_2$, we mix two anchor segments $s_1$, $s_2$ with data augmentation to constitute a mixture $x = s_1 + s_2$. Then the mixture is fed into the query-based source separator, described by the following regression:

\begin{equation}
  \label{f:separator}
  f(x, e_j) \mapsto s_j, j \in \{1, 2\}
\end{equation}

\noindent Equation \eqref{f:separator} shows that the separated sound depends on both the input mixture and the latent source embedding. The latent source embedding provides information on which source is to be separated. 

We follow the same setup as described in \cite{kong2023universal} and utilize a residual UNet30 (ResUNet30) to construct our source separator. The ResUNet30 is composed of 6 encoder blocks, 1 bottleneck block, and 6 decoder blocks. Each encoder block includes a single residual convolutional block to downsample the audio spectrogram into a bottleneck feature map. Each decoder comprises a single residual deconvolutional block to upsample the feature and obtain the individual sources. A skip connection is established from each encoder block to the corresponding decoder block with the same downsampling/upsampling rate. For a more detailed model architecture, refer to \cite{kong2021decoupling}. The latent source embedding is incorporated into the ResUNet separator using feature-wise linear modulation (FiLM) \cite{perez2018film} method.
The separation network predicts a complex ideal ratio mask (IRM), which can then be multiplied by the STFT of the mixture to derive the STFT of the separated source. By applying the inverse STFT (iSTFT), the waveform of the separated source can be obtained.
\subsection{Proposed Approach}\label{PA}
A-MAE originates from the image masked autoencoder (MAE) and learns self-supervised representations from Mel-spectrograms \cite{huang2022masked}. The architecture of A-MAE consists of a 12-layer Vision Transformers-Base (ViT-B) encoder, followed by a decoder comprising 8 Transformer blocks. The encoder of A-MAE partitions the spectrogram of each 10-second AudioSet recording into non-overlapping grid patches, where each patch comprises 16-by-16 time-mel bins. Subsequently, features are extracted based on the sequential positions of these patches in the Mel-spectrogram, ultimately resulting in a 768-dimensional embedding sequence with a length of 512. Eighty percent of the spectrogram patches are randomly masked, and the remaining non-masked patches are used to reconstruct the input spectrogram.

We propose using a pre-trained A-MAE encoder as an upstream model to extract universal features. In downstream USS tasks, the original decoder is discarded and replaced with ResUNet. The ResUNet takes the concatenation of the mixture STFT magnitude and A-MAE encoder features as input, as depicted by the dashed box in Figure \ref{framework}. Since A-MAE is pre-trained on the full AudioSet training set through self-supervised learning, it is expected to enhance the performance of downstream separation tasks.



Additionally, in our baseline system, an energy-matching data augmentation strategy has been proven effective in enhancing separation performance \cite{kong2023universal}. However, we observed that mixture source pairs obtained from this strategy sometimes have amplitudes that are too large, leading to waveform distortion. Therefore, we use Equation \eqref{amp_norm} to normalize the amplitudes of the mixture source pairs \((x, s)\) obtained after energy matching\textcolor{black}{\cite{liu2023separate}}.

\begin{equation}
  \label{amp_norm}
  x = x / \max_i |x_i|, \quad s = s / \max_i |s_i|
\end{equation}

\section{Experiments}\label{Exper}
\subsection{Training Dataset and Training Details}\label{TD}
\subsubsection{Training Dataset}
AudioSet \cite{gemmeke2017audio} is a large-scale weakly labeled audio dataset consisting of 527 audio events, spanning a broad spectrum of human and animal sounds, musical instruments and genres, as well as everyday environmental sounds. The balanced subset consists of over 20,000 sound clips, upon which our USS system is trained. Weakly labeled data means having labels for what types of sounds are present in a sound clip, but without exact information about when these sound events occur. We preprocessed all 10-second audio samples by resampling them to 32 kHz and converting them to single-channel.

\subsubsection{Training Details}
The duration of each sound mixture and target source is 2 seconds. To compute the STFT, we utilize a \textcolor{black}{Hann} window with a window size of 1024 and a hop size of 320. Given a sampling rate of 32 kHz, this results in 100 frames per second, maintaining alignment with PANNs. Each 2-second sound mixture contains 200 time frames, and we pad them with 24 zero-frames ($T=224$). To reduce the length of the input sequence, we introduce an average-max pooling operation \cite{liu2023simple} for the features of the A-MAE encoder. The shape of the A-MAE encoder feature after pooling is (32, 768), where 32 is the length of sequence, 768 is the embedding dimension. Therefore, along the sequence dimension, we duplicate the pretraining features 7 times to enable their concatenation with the STFT features along the feature dimension.

The embedding layer of PANNs has a dimension of 2048. Throughout the training of the USS system, the parameters of PANNs are frozen. The $l_1$ loss between the predicted separated source and the ground truth target source is used for training the query-based USS system. We adopt an Adam optimizer \cite{kingma2014adam} with a learning rate of 0.001. The entire training process consists of 60 epochs, with 10,000 iterations per epoch. The batch size is 16.

\subsection{Evaluation Dataset and Evaluation Methods}\label{EDEM}
\subsubsection{Evaluation Dataset}

The evaluation set of AudioSet consists of 18,887 sound clips with 527 sound events. The creation of the evaluation data follows the same process as that of the training data. \textcolor{black}{First, we use a pre-trained SED model to extract target anchor segments from 10-second audio clips. Then, we select anchor segments from two different sound classes and combine them to create a mixture.} For each sound class, we create 100 2-second mixtures to evaluate the separation results. In total, there are 52,700 evaluation pairs. In a similar manner, we create 52,700 mixture source pairs from the AudioSet balanced subset for the calculation of average embedding during the inference stage.

\subsubsection{Latent Source Embedding Calculation}

For query-based USS, there are two types of latent source embeddings used to evaluate the performance of the separation system. One is the oracle (ideal) embedding, calculated as in the training process, using the ground truth of the target source.  This can be expressed with the formula:

\begin{equation}
  \label{f_emb}
  e = f_{\text{LSE}}(s)
\end{equation}
where $s$ is the target source signal, and $f_{\text{LSE}}$ is the latent source embedding processor. Using oracle embeddings as the query condition reflects the maximum performance value for a general sound separation system.

In the inference process or practical applications, the ground truth of the target source is unknown, and we can only approximate it by collecting $N$ clean clips of the target source. The embedding obtained by calculating the mean of latent embeddings for these $N$ segments is referred to as the average embedding. In the evaluation, the average embedding is calculated by:

\begin{equation}
  \label{avg_emb}
  e = \frac{1}{N} \sum_{n=1}^{N} f_{\text{LSE}}(s_n)
\end{equation}
where $\{s_n\}_{n=1}^{N}$ are audio samples of the queried sound class, and $N$ represents the total number of these audio samples.

\subsubsection{Evaluation Metrics}

Following the previous works such as \cite{kong2023universal, chen2022zero}, we use signal-to-distortion ratio (SDR) and  signal-to-distortion ratio improvement (SDRi) as metrics to evaluate the performance of the USS system. A higher value indicates better separation performance.

\section{Results and Discussion}\label{RD}
We evaluate our proposed method\textcolor{black}{s} and the baseline on the evaluation dataset in Section \ref{EDEM}. To ensure a fair comparison, we have reproduced the results of the baseline system\footnote{\url{https://github.com/bytedance/uss}} to serve as our reference for comparison. 


\begin{table}[!ht]
\centering
\caption{The experimental results of comparing the separation performance of different models.}
\label{tab1}
\setlength{\tabcolsep}{2mm}{
\begin{tabular}{@{}cccc@{}}
\bottomrule
& \multicolumn{1}{c}{ora. emb} & \multicolumn{2}{c}{avg. emb} \\
\cmidrule(lr){2-2} \cmidrule(lr){3-4}
& SDRi (dB) & SDR (dB) & SDRi (dB) \\
\midrule
ResUNet \cite{kong2023universal} & 7.90 & 5.23 & 5.18 \\
MAE-ResUNet (frozen) & 8.45 & 5.53 & 5.62 \\
MAE-ResUNet (updated) & 8.45 & 5.57 & 5.64 \\
\bottomrule
\end{tabular}}
\end{table}

Table \ref{tab1} shows a comparison of the SOTA method \cite{kong2023universal} and our proposed methods on the evaluation data created using AudioSet, with results for SDR and SDRi. In the proposed methods, ``frozen'' indicates that during the fine-tuning process, the parameters of the A-MAE  are frozen, while ``updated'' signifies that we compute a weighted sum of the outputs from all A-MAE transformer layers and update the parameters of these layers. The average embedding (avg. emb) shows the results using Equation \eqref{avg_emb} as a condition. Compared to the oracle embedding (ora. emb), the use of average embedding better reflects the performance of the USS system in real-world applications.

From the table, it can be observed that our frozen method achieved an AudioSet SDRi of 5.62 dB using the average embedding, surpassing the SOTA system's 5.18 dB by 0.44 dB. This indicates that the use of universal features extracted by the self-supervised pre-trained model A-MAE and our data augmentation strategy contributes to the enhancement of the USS system's performance. However, compared to the frozen method, the effects of our updated method are not significant. Perhaps there are better updating methods worthy of further exploration.

\begin{figure}[!ht]
  \centering
  \includegraphics[scale=0.54]{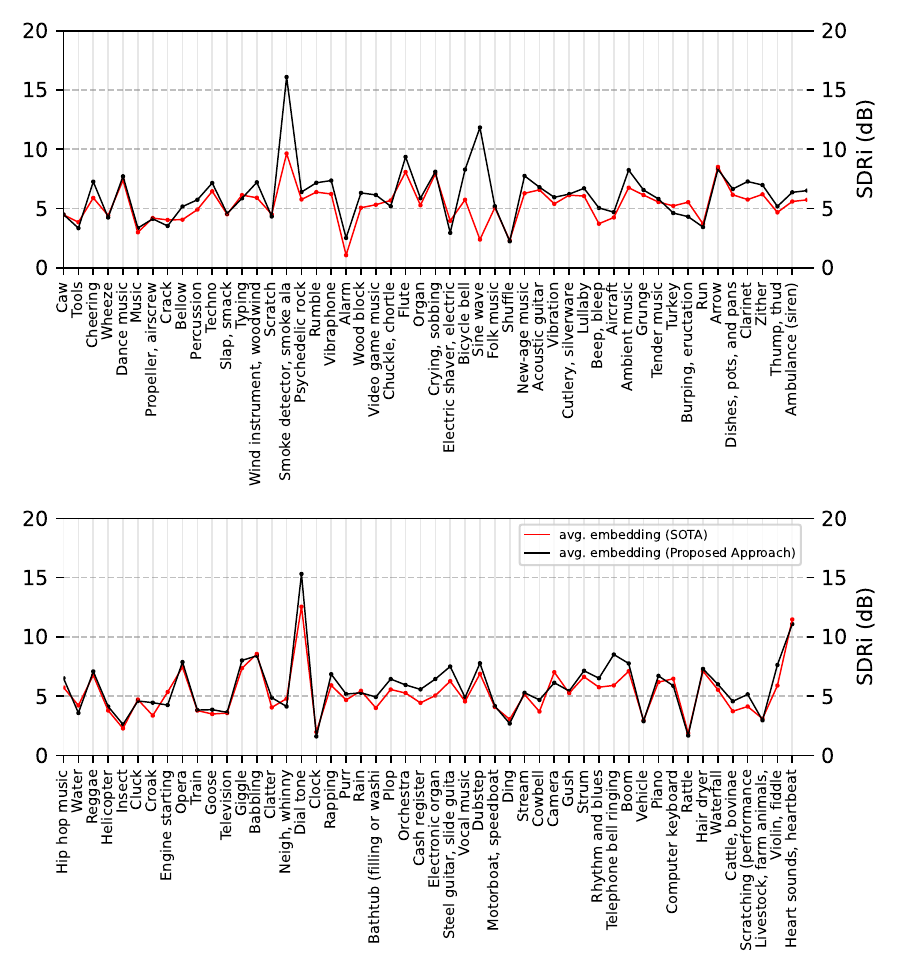}
  \caption{Class-wise USS results on some AudioSet sound classes.}
  \label{ch3:classwise_sep}  
\end{figure}


To further compare our frozen approach with the SOTA method, we plotted the class-wise SDRi results of AudioSet separation for 100 sound classes out of 527, as shown in Figure \ref{ch3:classwise_sep}. The red line illustrates the SDRi of the SOTA method using the average embedding. The black line represents the SDRi of our proposed method under the condition of average embedding. We observed that the black curve typically lies above the red curve, indicating an overall performance improvement of our proposed method over the SOTA method in most sound classes. Figure \ref{ch3:classwise_sep} indicates that our proposed method achieved an SDRi of over 15 dB in some sound classes, such as dial tone and smoke detector. All classes achieved positive SDRi scores. Compared to the SOTA method, certain sound classes, such as sine wave, smoke detector, and dial tone, exhibited the maximum improvement in SDRi. We discovered that these sounds share common line spectrum characteristics. However, for a very small number of sound classes, SDRi may not have improved or may have even worsened. \textcolor{black}{Additionally, for speech, our method achieves an SDRi of 5.86 dB.}

\begin{figure}[!ht]
  \centering
  \includegraphics[width=\columnwidth]{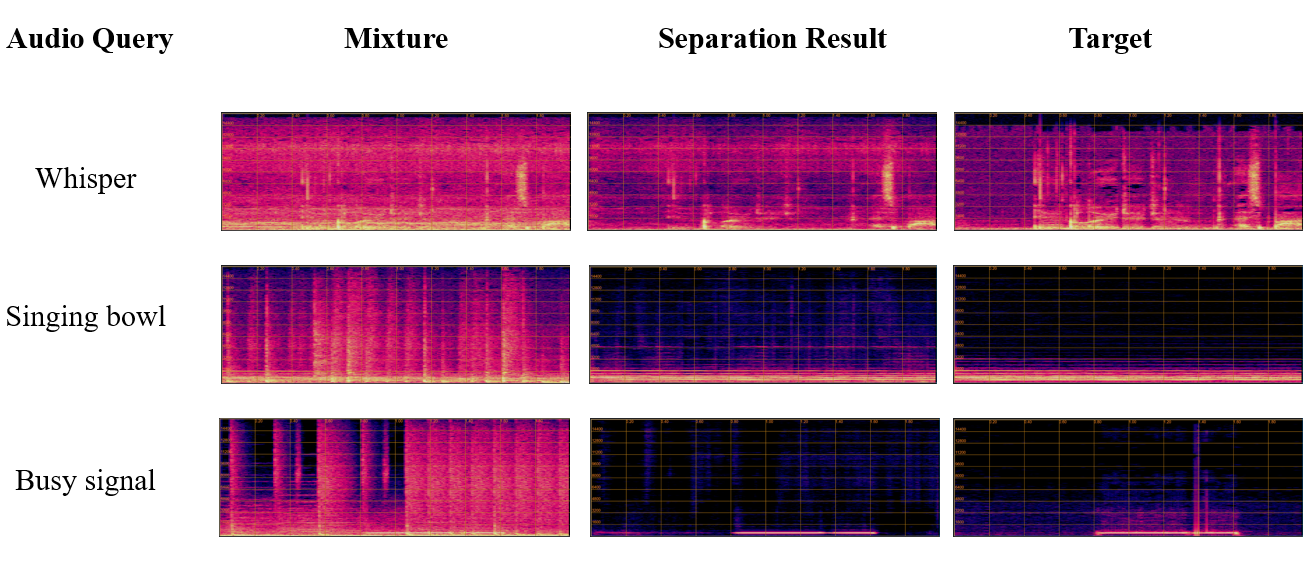}
  \caption{Visualization of separation results obtained by our model.}
  \label{visualization}  
\end{figure}
To demonstrate the separation performance of our proposed method, we visualize the spectrograms of the sound mixture, ground truth target sources, and sources separated using the average embedding of a specific sound class, as depicted in Figure \ref{visualization}. We noticed that the spectrogram pattern of the separated source closely resembles the ground truth of the target source, proving that our method is highly effective.




\section{Conclusion and Future Work}\label{Conclu}
For the first time, we proposed the application of self-supervised pre-trained audio MAE for universal sound separation tasks. By integrating STFT features and SSL embeddings, the performance of universal sound separation is improved. Experimental results indicate that, compared to the SOTA method, our proposed method achieves a 0.44 dB improvement in SDRi across 527 sound classes in AudioSet. In future work, we aim to enhance the separation ability of our system for unseen sound sources and plan to explore more modalities for universal sound separation.


\section*{Acknowledgment}
This work was supported in part by the Engineering and Physical Sciences Research Council (EPSRC) under Grant EP/T019751/1, funded by a PhD scholarship from the Doctoral College and the Centre for Vision, Speech, and Signal Processing (CVSSP) at the University of Surrey. For the purpose of open access, the authors have applied a Creative Commons Attribution (CC BY) licence to any Author Accepted Manuscript version arising.
\bibliographystyle{IEEEtran}
\bibliography{conference_101719}

\end{document}